\documentclass[aps,prl,reprint,superscriptaddress]{revtex4-2}
\usepackage{color}
\usepackage{graphicx}
\usepackage{amsmath}
\usepackage{bm}
\begin{document}

\title{Pressure-induced $\mathcal{PT}$ Symmetry Breaking in LaMnSi}

\author{Takuya Aoyama}
\email[]{aoyamatakuya@hiroshima-u.ac.jp}
\affiliation{Department of Quantum Matter, Graduate School of Advanced Science and Engineering, Hiroshima University, 1-3-1 Kagamiyama, Higashi-Hiroshima, Hiroshima 739-8530, Japan}
\affiliation{International Institute for Sustainability with Knotted Chiral Meta Matter (WPI-SKCM$^2$), Hiroshima University, 1-3-1 Kagamiyama, Higashi-Hiroshima, Hiroshima 739-8531, Japan}

\author{Hikaru Taneoka}
\affiliation{Department of Physics, Graduate School of Science, Tohoku University, 6-3 Aramaki-Aoba, Aoba-ku, Sendai, Miyagi 980-8578, Japan}

\author{Takemi Yamada}
\affiliation{Liberal Arts and Sciences, Toyama Prefectural University, Imizu, Toyama 939-0398, Japan}

\author{Hiroshi Tanida}
\affiliation{Liberal Arts and Sciences, Toyama Prefectural University, Imizu, Toyama 939-0398, Japan}

\author{Kenya Ohgushi}
\affiliation{Department of Physics, Graduate School of Science, Tohoku University, 6-3 Aramaki-Aoba, Aoba-ku, Sendai, Miyagi 980-8578, Japan}

\date{\today}

\begin{abstract}
We investigate the magnetotransport properties of the antiferromagnetic metal LaMnSi, in which the collinear magnetic order breaks both spatial inversion ($\mathcal{P})$ and time-reversal ($\mathcal{T}$) symmetry yet preserves their combined $\mathcal{PT}$ symmetry. 
High pressure is found to suppress this $\mathcal{PT}$-symmetric antiferromagnetic phase, inducing a transition into a $\mathcal{PT}$-broken state characterized by a large anomalous Hall effect. 
Based on symmetry analysis, we propose a candidate magnetic structure for the high-pressure phase. Subsequent band calculations for this structure reveal the emergence of band splitting and orbital-dependent spin polarization.
Our results establish LaMnSi as an ideal platform for controlling $\mathcal{PT}$ symmetry breaking via external parameters.

\end{abstract}

\maketitle

Electronic phases characterized by broken spatial inversion ($\mathcal{P}$) and time-reversal ($\mathcal{T}$) symmetries are of central importance in condensed matter physics~\cite{Fiebig2005, Spaldin2008,Tokura2014, Jungwirth2016, Baltz2018}.
In such systems, the presence or absence of the combined $\mathcal{PT}$ symmetry plays a crucial role.
If the $\mathcal{PT}$ symmetry is broken, electric and magnetic degrees of freedom can exhibit distinct ferroic orders.
Conversely, materials with preserved $\mathcal{PT}$ symmetry exhibit unique cross-correlations, such as the magnetoelectric effect in antiferromagnetic insulators~\cite{Dzyaloshinskii1960,Astrov1960,Rado1961}.
Recently, this concept has been extended to metals, within which the preserved $\mathcal{PT}$ symmetry stabilizes spin-degenerate bands with the $\varepsilon(\vec{k}) \neq \varepsilon(-\vec{k})$ characteristic~\cite{Yuan2021, Hayami2024}, leading to novel quantum phenomena such as current-induced N\'eel vector switching, magnetic field--induced strain, and nonlinear conduction~\cite{Zelezny2014, Wadley2016, Watanabe2017, Watanabe2020, Shiomi2019, Yatsushiro2021, Tanida2023}.
Beyond exploring novel functionalities in these metals, it is highly expected that novel quantum phases may be observed when suppressing the $\mathcal{PT}$-symmetric phase via parameter tuning~\cite{Gegenwart2008,Cao2023}.

\begin{figure}[th]
\centering
\includegraphics[width=8cm,pagebox=cropbox,clip]{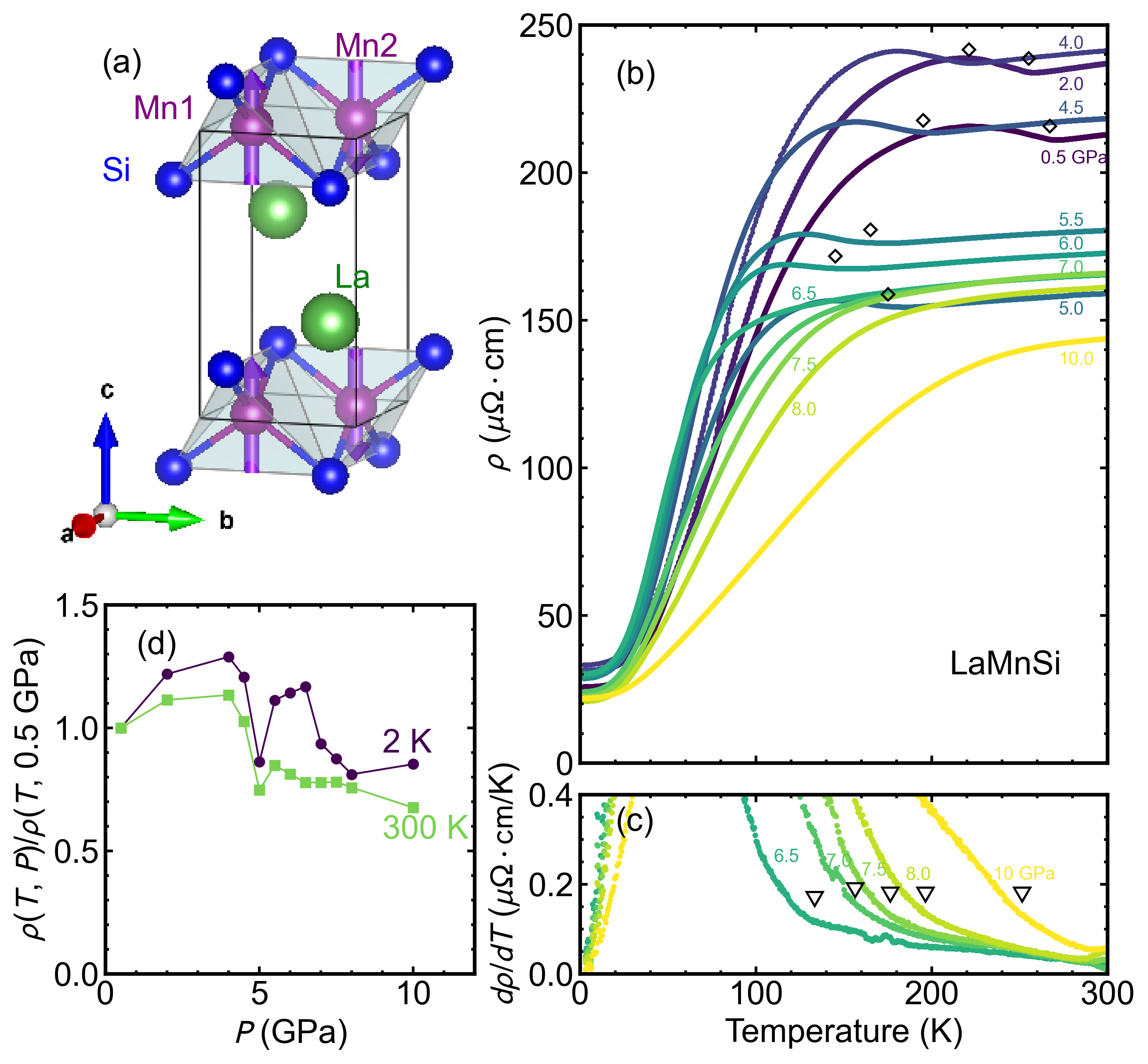}
\caption{(a) Crystal and magnetic structures of LaMnSi visualized using the VESTA tool~\cite{Momma2011}. Temperature ($T$) dependence of (b) the electrical resistivity ($\rho_{xx}$) and (c) $\frac{\rm d\rho}{{\rm d}T}$ above 6.5~GPa, under high pressures. Open diamonds and triangles indicate the antiferromagnetic and ferromagnetic transitions, respectively. (d) 
Pressure dependence of electrical resistivity scaled by the value at 0.5~GPa, $\rho$($T$, $P$)/$\rho$($T$, 0.5~GPa).}
\label{structure}
\end{figure}

\begin{figure*}[th]
\centering
\includegraphics[width=18cm,pagebox=cropbox,clip]{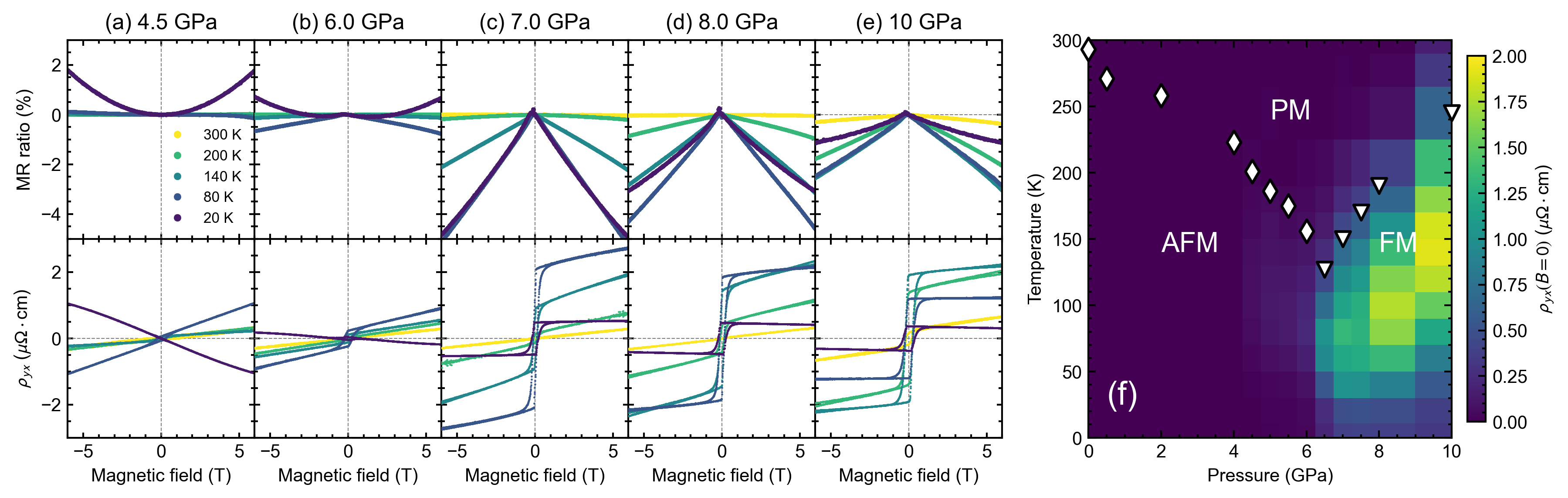}
\caption{Magnetoresistance (upper) and Hall resistivity (lower) at selected pressures of (a) 4.5, (b) 6.0, (c) 7.0, (d) 8.0, and (e) 10~GPa (Run 1).
The electric current was applied along the 100 axis, and the magnetic field was applied along the 001 axis.
The data were collected after field cooling with 9~T from room temperature. 
(f) Temperature--pressure phase diagram for LaMnSi. The open diamonds and triangles indicate $T_{\rm N}$ and $T_{\rm c}$, respectively. The Hall resistivity at $B$ = 0 is superimposed as a contour plot. PM, AFM, and FM indicate the paramagnetic, antiferromagnetic, and ferromagnetic states, respectively.
}
\label{Hall}
\end{figure*}

Our target material LaMnSi possesses a CeFeSi-type crystal structure (Fig.~\ref{structure}(a)), which belongs to the space group of $P4/nmm$~\cite{Welter1994a, Tanida2022}.
This space group preserves the $\mathcal{P}$ symmetry, whose origin is located at the midpoints of the two Mn atoms in a unit cell, corresponding to the $4d$ and $4e$ Wyckoff positions with a site symmetry of $2/m$.
Because the unit cell contains two formula units of LaMnSi and consequently an even number of electrons, it is expected to be either a band insulator or a compensated semimetal.
Experimentally, LaMnSi has been reported to exhibit metallic electrical conductivity~\cite{Tanida2022}.
The material shows an antiferromagnetic (AFM) order with the wave vector $\vec{Q}$ = 0 at a N\'{e}el temperature ($T_{\rm N}$) of 293~K.
In this magnetic order, Mn spins pointing to the $c$-axis order antiferromagnetically in the $ab$-plane and ferromagnetically along the $c$-axis, forming a C-type AFM structure~\cite{Sakai2026}.
According to the representation analysis (see the Supplemental Material (SM)), this magnetic structure corresponds to the $B_{1u}$ irreducible representation in the point group of $4/mmm$.
This AFM phase possesses the magnetic point group of $4^{\prime}/m^{\prime}m^{\prime}m$, in which the ${\mathcal P}$ symmetry is broken because the two Mn spins in a unit cell point in opposite directions and the ${\mathcal T}$ symmetry is broken as a result of the AFM order.
Nevertheless, the combined $\mathcal{PT}$ symmetry is still preserved because the ${\mathcal T}$ operation reverses the spin direction.
Hence, LaMnSi is a unique antiferromagnet with ${\mathcal P}$ odd, ${\mathcal T}$ odd, and $\mathcal{PT}$ even characteristics~\cite{Iwata2026}.
The magnetic order leads to a peculiar anomaly in the electrical resistivity: a small but clear upturn near $T_{\rm N}$.
This upturn suggests a partial opening of an energy gap at the Fermi energy.
This phenomenon is frequently attributed to a band folding due to the AFM order; however, this scenario is not applicable to LaMnSi with $\vec{Q}=0$.
Instead, it is argued to be related to the broken $n$-glide symmetry based on the magnetic order~\cite{Tanida2022}.
Namely, band degeneracies protected by the glide symmetry at the zone boundary above $T_{\rm N}$ are lifted by the magnetic order, leading to a larger electrical resistivity.

In this study, we investigate the electrical transport properties under high pressure and magnetic field for a ${\mathcal P}{\mathcal T}$ symmetric antiferromagnet LaMnSi with a metallic ground state.
By tracking the resistivity anomaly at $T_{\rm N}$, we show that the AFM order is suppressed by the application of pressure and completely disappears at approximately 5~GPa.
We also find that a ferromagnetic (FM) phase characterized by a finite anomalous Hall effect (AHE) emerges above 5~GPa.
We discuss the origin of this pressure-induced phase in terms of a localized-to-itinerant crossover driven by the contraction of the Mn--Mn interatomic distance, which leads to a symmetry-breaking transition from a ${\mathcal P}{\mathcal T}$-symmetric AFM state to a ${\mathcal P}{\mathcal T}$-broken FM state.

Single crystals of LaMnSi were grown via the self-flux method as described in Ref.~\cite{Tanida2023, Tanida2022}.
Electrical transport measurements under high pressure ($P$) were performed using a diamond anvil cell with NaCl as the pressure medium.
The pressure was calibrated using the ruby fluorescence method.
The resistivity ($\rho_{xx}$) and Hall resistivity ($\rho_{yx}$) were measured using a standard four-probe method with the electrical current $I \parallel [100]$ and magnetic field $B \parallel [001]$ under various temperatures from 2~K to 300~K.
We measured two samples (Run 1 and Run 2) to ensure reproducibility (see the SM).
The main text focuses on the results of Run 1, which cover a wider pressure range.
Computational details regarding the first-principles calculations are provided in the SM.

The temperature ($T$) dependence of the electrical resistivity of LaMnSi is shown in Fig.~\ref{structure}(b).
At 0.5~GPa, the resistivity exhibits metallic behavior, characterized by a decrease in $\rho_{xx}$ upon cooling.
A closer inspection from the high-temperature side reveals a local minimum in $\rho_{xx}$ at $T_{\rm N}$ = 272~K, which is attributed to the AFM order as indicated by the open diamonds.
With further cooling, $\rho_{xx}$ increases moderately down to 240~K and then decreases rapidly toward the base temperature.
The residual resistivity ratio is approximately 17, indicating high crystal quality.

Applying pressure up to 6~GPa suppresses the upturn associated with the AFM transition down to 150~K.
Above 6~GPa, the temperature dependence of the resistivity exhibits a qualitative change: the distinct upturn associated with the AFM order disappears and the resistivity decreases monotonically upon cooling.
A detailed analysis of the differential resistivity $\frac{d\rho}{dT}$ reveals a sharp decrease in $\rho$ below a characteristic temperature indicated by the open triangles in Fig.~\ref{structure}(c).
A qualitative change in this resistivity anomaly under applied pressure has also been reported in CeMnSi~\cite{Nishiyama2026}.
This temperature turns out to be an FM transition temperature $T_{\rm c}$, below which the AHE emerges.
The collected $T_{\rm N}$ and $T_{\rm c}$ values allow us to map out the temperature--pressure phase diagram (Fig~\ref{Hall}). 

As shown in Fig.~\ref{structure}(d), the pressure dependence of the electrical resistivity provides crucial insights into the changes in the electronic state associated with the pressure application.
As the pressure increases up to 4~GPa, the resistivity gradually increases, likely reflecting enhanced scattering from spin fluctuations.
Over a broad pressure range from 4.5~GPa to 6~GPa, both the residual resistivity ($\rho_{\mathrm{2K}}$) and the room-temperature resistivity ($\rho_{\mathrm{300K}}$) exhibit a clear drop.
Such a significant suppression even above the magnetic transition temperature suggests a Fermi surface reconstruction across this phase boundary.

To investigate the details of the pressure-induced phase transition, we measured the magnetoresistance (MR) and Hall resistivity under high pressure.
Figure~\ref{Hall} displays the magnetic field ($B$) dependence of the MR and Hall resistivity at selected temperatures and pressures.
At 4.5~GPa (Fig.~\ref{Hall}(a)), a positive MR proportional to $B^2$ is observed and its amplitude increases upon cooling, which is consistent with normal MR behavior.
This behavior supports the simple expectation that LaMnSi is a compensated semimetal.
Indeed, as shown in the SM, the magnetotransport properties below 4.5~GPa are well described by a two-carrier Drude model: the estimated carrier density and mobilities are $n \approx 2 \times 10^{20}$~cm$^{-3}$ and $\mu_{\mathrm{e}} \sim \mu_{\mathrm{h}} \sim 400$~cm$^2$/Vs, respectively.
Upon further pressurization up to 6~GPa (Fig.~\ref{Hall}(b)), a new component of negative MR emerges below 140~K.
This is most likely caused by the suppression of scattering due to spin fluctuations by the application of $B$ in the FM state.
Indeed, at 140~K, where negative MR appears, hysteresis is observed in $\rho_{yx}$ below $\pm 1$~T, strongly indicating that the anomalous Hall resistivity $\rho_{yx}$ is associated with the FM order.
As the pressure is further increased to 7~GPa (Fig.~\ref{Hall}(c)), the negative MR and AHE become more pronounced.
With increasing pressure to 8~GPa, the negative MR is weakened and the temperature at which negative MR and the AHE is most prominent shifts to higher temperatures (Fig.~\ref{Hall}(d)).
This trend continues to 10~GPa, where the negative MR is weakened moderately and the hysteresis in $\rho_{yx}$ becomes larger (Fig.~\ref{Hall}(e)).
These observations indicate that the FM order stabilizes with increasing pressure.

\begin{figure}[th]
\centering
\includegraphics[width=8cm,pagebox=cropbox,clip]{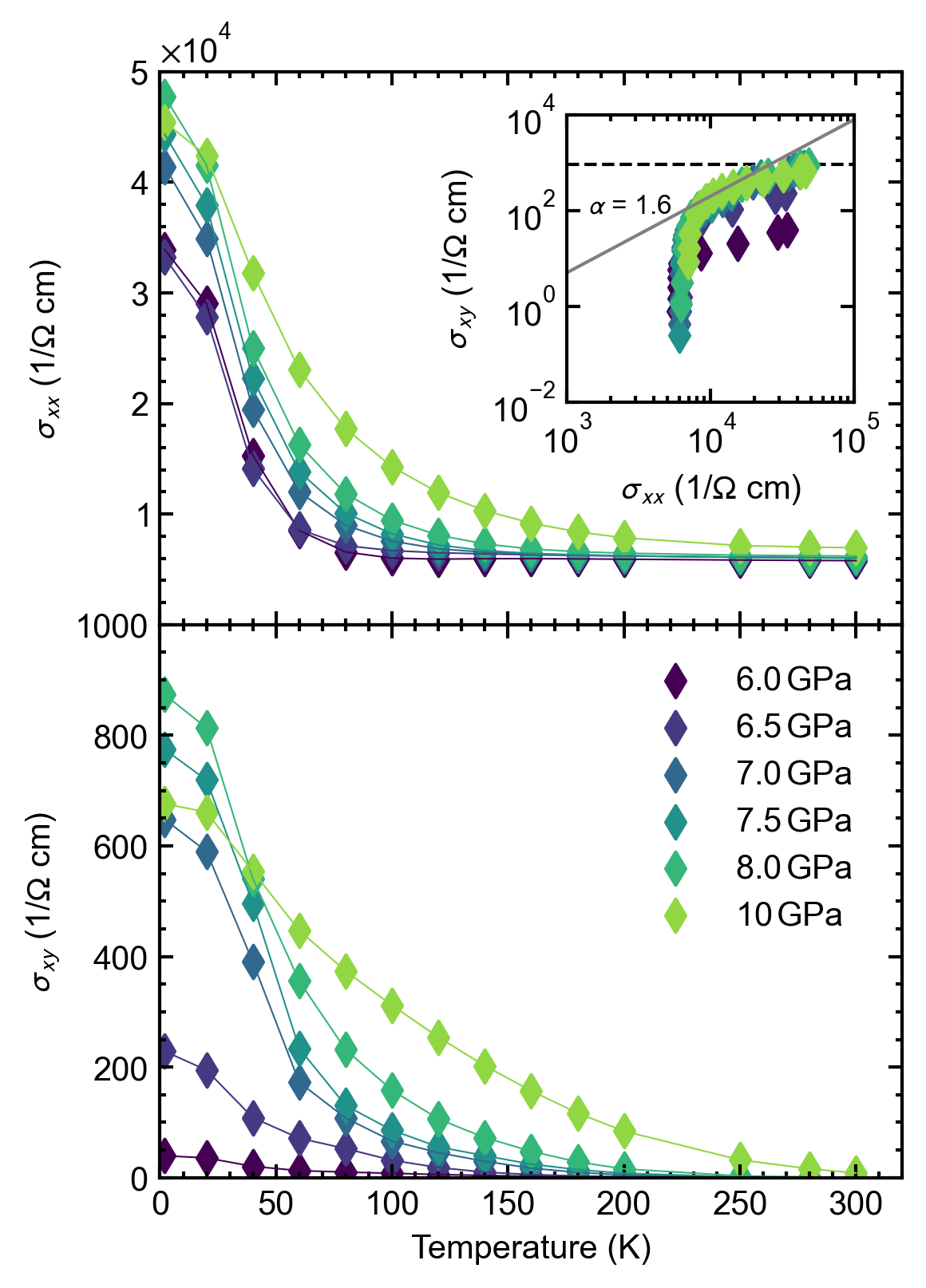}
\caption{Temperature dependence of the electrical conductivity and Hall conductivity collected under several pressures. The inset shows a plot of $\sigma_{xx}$ versus $\sigma_{xy}$. The gray lines indicate the power law of $\sigma_{xy} \propto \sigma_{xx}^{\alpha}$, with $\alpha = 1.6$. The horizontal broken line indicates $\sigma_{xy} = e^2/(ha) = 926~\Omega^{-1}$cm$^{-1}$, with $a = 4.18$~\AA}
\label{sigma}
\end{figure}

We performed the experiments described in the previous paragraph at various temperatures and pressures. 
The collected results are superimposed on the phase diagram as a contour map (Fig.~\ref{Hall}(f)).
We then calculated the electrical and Hall conductivities at zero magnetic field, $\sigma_{xx} = \rho_{xx}/(\rho_{xx}^2 + \rho_{yx}^2)$ and $\sigma_{xy} = \rho_{yx}/(\rho_{xx}^2 + \rho_{yx}^2)$, which are plotted in Fig.~\ref{sigma}.
On cooling, $\sigma_{xx}$ shows an increase reflecting metallic conduction, while $\sigma_{xy}$ appears around the FM transition temperature.
Both $\sigma_{xx}$ and $\sigma_{xy}$ tend to saturate at low temperatures.
Pressurization results in an increase in the magnitudes of $\sigma_{xx}$ and $\sigma_{xy}$, and the onset temperature of the AHE shifts to higher temperatures.
Typical values are $\sigma_{xx} \sim 10^4~\Omega^{-1}$cm$^{-1}$, while the anomalous Hall conductivity ranged from $\sigma_{xy} = 1~\Omega^{-1}$cm$^{-1}$ to $10^3~\Omega^{-1}$cm$^{-1}$.
Note that the maximum $\sigma_{xy}$ value is comparable to that observed in FM metals such as Fe~\cite{Dheer1967,Yao2004}. 

The mechanism of the AHE is classified into three regimes based on the scattering strength ~\cite{Onoda2008,Nagaosa2010,Miyasato2007}.
In the dirty/hopping regime of $\hbar/\tau \gtrsim E_{\rm F}$ ($\tau$ and $E_{\rm F}$ being the transport relaxation time and the Fermi energy, respectively), a scaling relation of $\sigma_{xy} \propto \sigma_{xx}^{\alpha}$ with an exponent of $\alpha \ge 1.6$ is empirically known.
In the intermediate regime of $\varepsilon_{\rm SO} < \hbar/\tau < E_{\rm F}$ ($\varepsilon_{\rm SO}$ being the spin--orbit coupling energy), $\sigma_{xy}$ is primarily governed by the intrinsic Berry curvature, taking a nearly universal value of $\sim e^2/ha$, independent of $\sigma_{xx}$.
In the superclean limit of $\hbar/\tau \ll \varepsilon_{\rm SO}$, skew scattering prevails, leading to a linear relationship of $\sigma_{xy} \propto \sigma_{xx}$.
The inset of Fig.~\ref{sigma} illustrates the $\sigma_{xy}$ versus $\sigma_{xx}$ plot at $P \geq 6$~GPa, revealing a systematic evolution across these different transport regimes. 
In the high-temperature (low-conductivity) region, $\sigma_{xy}$ decreases steeply with $\sigma_{xx}$, following a power law of $\sigma_{xy} \propto \sigma_{xx}^{\alpha}$, with $\alpha \gtrsim 1.6$. 
This steep reduction is characteristic of the dirty/hopping regime, where the intrinsic Berry curvature contribution is significantly suppressed by strong scattering or carrier localization.
As the temperature decreases and $\sigma_{xx}$ increases, the scaling behavior smoothly connects to the $\alpha \approx 1.6$ slope. 
Finally, in the high-conductivity limit at low temperatures, $\sigma_{xy}$ tends to saturate and asymptotically approaches the quasi-universal value of $e^2/(ha)$. 
This saturation signifies the crossover into the intermediate regime, where the AHE is dominated by the intrinsic Berry curvature of the electronic bands rather than impurity scattering~\cite{Onoda2008,Nagaosa2010}.

\begin{figure}[h]
\centering
\includegraphics[width=8.5cm,pagebox=cropbox,clip]{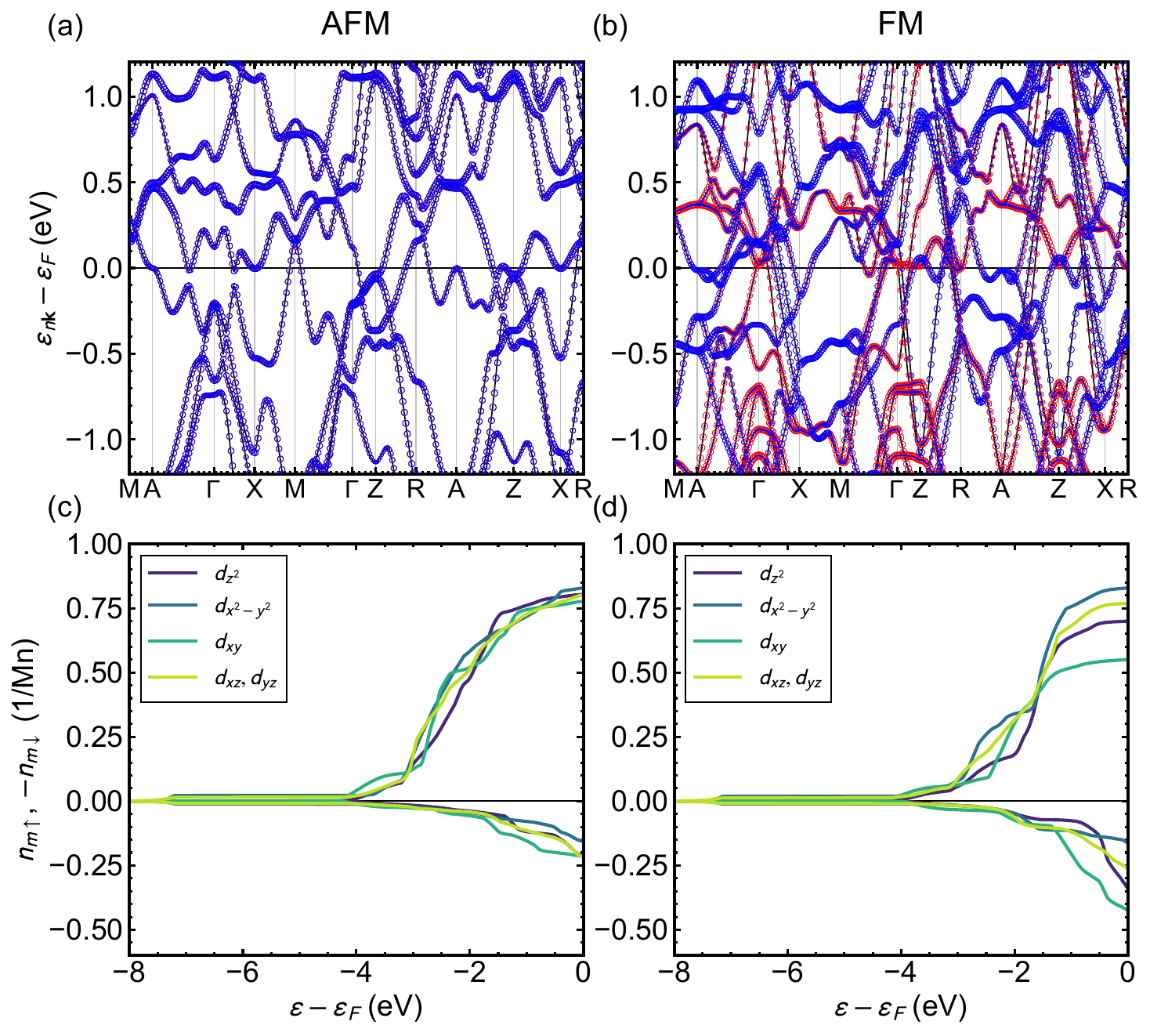}
\caption{Band structures (upper) and energy-dependent spin orbital occupancy of Mn $3d$ states below the Fermi level (lower) for LaMnSi. In the band structures, the up- and down-spin components of the Mn1 atom are represented by the sizes of the red and blue circles, respectively. Only the Mn1 atom is shown because the relations Mn2$\uparrow$ = Mn1$\uparrow$ and Mn2$\downarrow$ = Mn1$\downarrow$ hold in the FM state, whereas Mn2$\uparrow$ = Mn1$\downarrow$ and Mn2$\downarrow$ = Mn1$\uparrow$ hold in the AFM state. Calculations assume the (a, c) AFM ($B_{1u}$) and (b, d) FM ($A_{2g}$) states.
}
\label{band}
\end{figure}

To gain microscopic insight into the experimentally observed electronic reconstruction and anomalous transport, we performed first-principles calculations based on density functional theory for both the experimentally established $B_{1u}$ AFM state and a candidate $A_{2g}$ FM state, as discussed in the SM (Fig.~\ref{band}(a) and \ref{band}(b)).
In the AFM phase, the $\mathcal{PT}$ symmetry is preserved, enforcing double degeneracy of the electronic bands at each $\vec{k}$ point.
The resulting band structure exhibits characteristic warping, as detailed in the SM.
Conversely, in the candidate FM phase, the breaking of time-reversal symmetry lifts the $\mathcal{PT}$ protection, leading to a breaking of the band degeneracy associated with spin polarization.
Such symmetry lowering provides consistent results for the observed AHE.

Beyond symmetry considerations, our calculations indicate an orbital-dependent redistribution of spin-resolved occupations within the Mn $3d$ manifold as a possible microscopic origin of the transition.
Figure~\ref{band}(c) and \ref{band}(d) shows the spin-resolved occupation of Mn1 as a function of energy for the AFM and FM phases, respectively, where $n_{m\sigma}$ denotes the occupation of orbital $m$ with spin $\sigma=\uparrow,\downarrow$.
In the AFM phase, the spectral weight is primarily located at higher binding energies with nearly uniform orbital occupations and the magnetic moment is $3.08~\mu_{\mathrm{B}}$. 
Conversely, in the FM phase, the spectral weight is redistributed toward the Fermi level, indicating enhanced itinerancy.
This redistribution is accompanied by an increased minority-spin contribution and a reduction of the net spin polarization; this is particularly pronounced for the $d_{xy}$ orbital.
Accordingly, the magnetic moment is reduced to $2.20~\mu_{\mathrm{B}}$.
In this more itinerant regime, the kinetic energy gain of delocalized electrons, together with Hund's coupling, provides a plausible mechanism for stabilizing the FM state under pressure.
The enhanced itinerancy in such an FM state is consistent with a smaller $\rho_{xx}(T)$ compared with that in the ambient-pressure phase.

Finally, we discuss the structural origin of the transition from the $\mathcal{PT}$-symmetric AFM phase to the $\mathcal{PT}$-breaking FM phase.
Previous X-ray diffraction measurements under pressure have shown that the in-plane Mn--Mn distance ($d_{\mathrm{Mn-Mn}}$) decreases continuously from 2.95~{\AA} at 0~GPa to $\sim 2.83$~{\AA} at 10~GPa~\cite{Kawamura2026}.
Studies on chemically pressurized La$_{1-x}$Y$_x$MnSi systems have reported a corresponding transition from the AFM to FM order as $d_{\mathrm{Mn-Mn}}$ approaches a critical value of $\sim 2.83$~{\AA}~\cite{Ijjaali1998,Welter1994a}.
The qualitative consistency between the physical and chemical pressure effects suggests that the reduction of $d_{\mathrm{Mn-Mn}}$ is a key contributing factor driving the system toward the FM state.
For a more quantitative discussion, however, it is necessary to consider the three-dimensional structural modifications, including changes in the Mn--Si network and the resultant inter-orbital hybridizations.
Further diffraction experiments under high pressure will be necessary to clarify the microscopic mechanism of this transition.

In summary, we investigated the pressure-dependent magnetotransport properties of LaMnSi.
The ambient-pressure AFM phase with preserved $\mathcal{PT}$ symmetry is suppressed under pressure, and a new phase emerges above approximately 5~GPa, accompanied by a large AHE.
Our first-principles calculations suggest that a transition to an FM state lifts the $\mathcal{PT}$ protection and allows breaking of the band degeneracy due to spin polarization.
Furthermore, a comparison of the calculated electronic structures between the AFM and FM phases indicates the development of orbital-dependent spin polarization within the Mn $3d$ manifold.
Together with comparisons with chemically pressurized systems, our results suggest that the reduction of the in-plane Mn--Mn distance plays an important role in driving this phase transition.

\begin{acknowledgments}
    We thank K. Kuroda, Y. Kawamura, and Y. Imai for fruitful discussions. 
    This work is financially supported by JSPS KAKENHI Nos.
    JP25K22015, 
    JP25K00960, 
    JP25K00955, 
    JP25H01246, 
    JP25H01247, 
    JP24K06943, 
    and JP25K07209 
    and by the Murata Science Foundation.
\end{acknowledgments}

\bibliographystyle{apsrev4-2}
\bibliography{manuscript}

\clearpage
\appendix 
\setcounter{figure}{0}
\setcounter{table}{0}
\setcounter{equation}{0}
\setcounter{page}{1} 

\renewcommand{\thefigure}{S\arabic{figure}}
\renewcommand{\thetable}{S\arabic{table}}
\renewcommand{\theequation}{S\arabic{equation}}

\onecolumngrid 
\begin{center}
  \textbf{\large Supplemental Material: \\ 
  Pressure-induced $\mathcal{PT}$ Symmetry Breaking in LaMnSi} \\[.5cm]
  Takuya Aoyama, Hikaru Taneoka, Takemi Yamada, Hiroshi Tanida, and Kenya Ohgushi.
\end{center}

\section{Magnetic Symmetry Analysis}
Because experimental studies indicate that the crystal structure remains unchanged even under pressure~\cite{Kawamura2026}, we estimate the magnetic structure under pressure by assuming a magnetic propagation vector ($\vec{Q}$) based on the ambient-pressure crystal structure with $P4/nmm$ symmetry.

We first consider the case of $\vec{Q}$ = 0. 
The magnetic representation $\Gamma_{\rm mag}$ on the Mn site decomposes into the irreducible representations (IRs) such that 
\begin{equation}
\Gamma_{\rm mag} = A_{2g} + E_g + B_{1u} + E_u.
\end{equation}
Here, the even-parity representations $A_{2g}$ and $E_g$ correspond to ferromagnetic (FM) modes with moments aligned along the $c$-axis and in the $ab$-plane, respectively.
Conversely, $B_{1u}$ and $E_u$ are odd-parity representations describing antiferromagnetic (AFM) arrangements with zero net magnetization.
The magnetic structure of the ambient pressure phase is assigned to the $B_{1u}$ representation, a state that preserves $\mathcal{PT}$ symmetry (combined space inversion $\mathcal{P}$ and time-reversal $\mathcal{T}$ symmetry) and thus forbids emergence of the anomalous Hall effect.

The emergence of the AHE in the high-pressure phase strongly suggests a symmetry-breaking transition to a state with broken $\mathcal{PT}$ symmetry, such as the FM configurations allowed by the $A_{2g}$ or $E_g$ representations. 
Given that the AHE is observed with the magnetic field applied along the $c$-axis, we can conclude that the $A_{2g}$ representation, which allows for an FM component $F_z$, is the most plausible candidate for the high-pressure phase.
From a symmetry perspective, the $B_{1u}$ AFM state and the $A_{2g}$ FM state correspond to the magnetic point groups $4'/m'm'm$ and $4/mm'm'$, respectively.

\begin{table}[h]
\centering
\caption{\label{tab:irrep} Magnetic representation $\Gamma_{\mathrm{mag}}$ decomposed into irreducible representations (IRs) for the Mn site in LaMnSi ($\vec{Q} = 0$). $\mathrm{F}$ and $\mathrm{AF}$ denote the ferromagnetic and antiferromagnetic components, respectively.}

\begin{minipage}{0.4\columnwidth} 
\begin{ruledtabular}
\begin{tabular}{cccc}
 IR & $m_x$ & $m_y$ & $m_z$ \\
\hline
 $A_{2g}$ & & & $\mathrm{F}_z$ \\
 $E_{g}$  & $\mathrm{F}_x$ & $\mathrm{F}_y$ & \\
 $B_{1u}$ & & & $\mathrm{AF}_z$ \\
 $E_{u}$  & $\mathrm{AF}_x$ & $\mathrm{AF}_y$ & \\
\end{tabular}
\end{ruledtabular}
\end{minipage}
\end{table}

Next, we consider $\vec{Q} = (0, 0, 1/2)$, which is realized experimentally in PrMnSi and NdMnSi~\cite{Welter1994a}.
The results of the decomposition into IRs are shown in Table S2.
The superscripts $(+)$ and $(-)$ denote parallel and antiparallel alignments, respectively, between the two Mn atoms within the same unit cell.
The resulting magnetic structures can be viewed as stacking $\vec{Q} = 0$--like configurations along the $c$-axis with alternating phases.
Consequently, the magnetic unit cell is doubled along the $c$-axis.
Within each single irreducible representation, the magnetic structures are macroscopically antiferromagnetic and do not allow a uniform magnetization.
More generally, none of these IRs support an order parameter that transforms in the same way as the magnetic dipole moment.
This indicates that the experimentally observed AHE cannot be accounted for by any of these single-irrep magnetic structures alone~\cite{Suzuki2017}.

\begin{table}[h]
\centering
\caption{\label{tab:irrep_qhalf} Magnetic representation $\Gamma_{\mathrm{mag}}$ decomposed into IRs, labeled by the Mulliken symbols corresponding to the little co-group, for the Mn site in LaMnSi at the propagation vector $\vec{Q} = (0, 0, 1/2)$. Because of the modulation along the $c$-axis, all configurations represent macroscopically antiferromagnetic (AF) states. The superscripts $(+)$ and $(-)$ denote parallel and antiparallel spin alignments, respectively, between the two Mn atoms within the same unit cell.}

\begin{minipage}{0.45\columnwidth} 
\begin{ruledtabular}
\begin{tabular}{cccc}
 IR & $m_x$ & $m_y$ & $m_z$ \\
\hline
 $A_{2g}$ & & & $\mathrm{AF}^{(+)}_z$ \\
 $E_{g}$  & $\mathrm{AF}^{(+)}_x$ & $\mathrm{AF}^{(+)}_y$ & \\
 $B_{1u}$ & & & $\mathrm{AF}^{(-)}_z$ \\
 $E_{u}$  & $\mathrm{AF}^{(-)}_x$ & $\mathrm{AF}^{(-)}_y$ & \\
\end{tabular}
\end{ruledtabular}
\end{minipage}
\end{table}
\clearpage

\section{Magnetotransport properties obtained in Run 2}

\begin{figure}[h]
\centering
\includegraphics[width=8cm,pagebox=cropbox,clip]{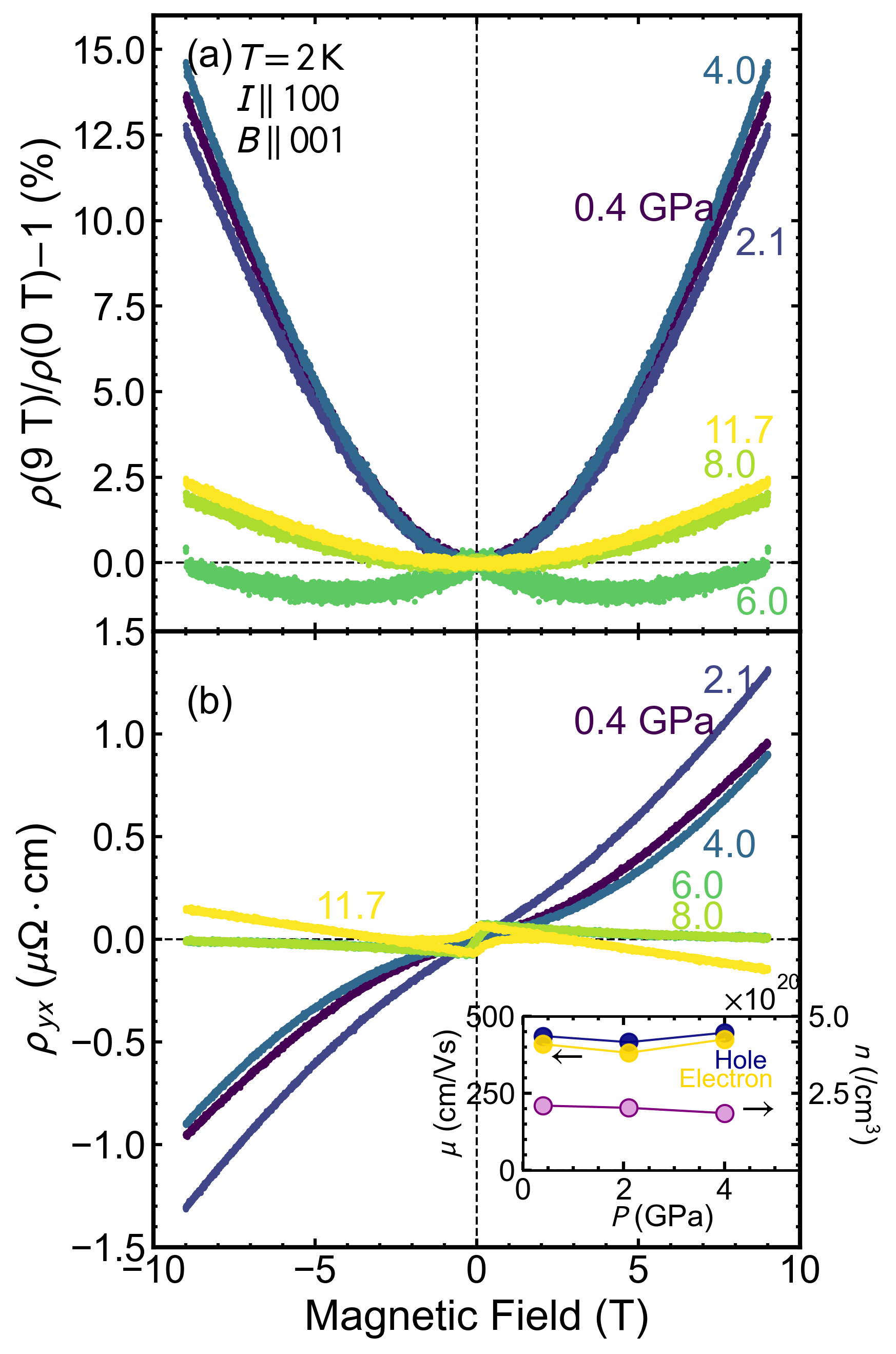}
\caption{(a) Magnetoresistance and (b) Hall resistivity under high pressures collected at 2~K in Run 2.}
\label{SFig1}
\end{figure}

In Run 2, we examined the magnetoresistance (MR) and Hall resistivity at a base temperature of 2~K under pressures up to 11.7~GPa.
The experimental set up was the same as in Run 1.
The data from Run 2 are shown in Fig.~\ref{SFig1}; these data are completely consistent with those of Run 1.
The MR at 0.4 GPa is proportional to $B^2$, which can be interpreted as the normal MR observed in a multi-carrier system.
The Hall resistivity at 0.4~GPa has not only a B-linear contribution with a hole characteristic but also a $B^3$ contribution.
This phenomenon is a typical feature of a multi-carrier system and is consistent with the observation of the normal MR.
The MR and Hall resistivity showed no significant changes under pressure up to 4~GPa, below which the $\mathcal{PT}$ symmetric AFM phase is the ground state.

The stable behavior of the $\mathcal{PT}$ symmetric phase observed at low pressures undergoes a dramatic transformation above 5~GPa.
The higher pressure regime exhibits a distinct AHE characterized by a rapid low-field increase in the Hall resistivity followed by clear saturation.
The AHE signal becomes increasingly pronounced with further compression, and the saturation field shifts toward lower values, reproducing the quantitative trends observed in Run 1.
Simultaneously, the MR profile transits from a simple $B^2$ dependence to complex behavior with a sharp low-field dip, reflecting the development of FM magnetization.
These consistent observations across independent experimental runs confirm the robust nature of the pressure-induced $\mathcal{PT}$ symmetry breaking in LaMnSi.

Here, we analyze the MR and Hall resistivity below 4.0~GPa by applying the Drude model for multi-band carriers. In the two-carrier model, the Hall resistivity $\rho_{yx}(B)$ is expressed as a power series of the magnetic field $B$:
\begin{equation}
\rho_{yx}(B) = R_1 B + R_3 B^3 + \mathcal{O}(B^5),
\end{equation}
where the coefficients $R_1$ and $R_3$ are given by
\begin{equation}
R_1 = \frac{q_1 n_1 \mu_1^2 + q_2 n_2 \mu_2^2}{(q_1 n_1 \mu_1 + q_2 n_2 \mu_2)^2}
\end{equation}
and
\begin{equation}
R_3 = -\frac{q_1 q_2 n_1 n_2 \mu_1^2 \mu_2^2 (q_1 n_1 + q_2 n_2)(\mu_1 - \mu_2)^2}{(q_1 n_1 \mu_1 + q_2 n_2 \mu_2)^4}.
\end{equation}
Here, $q_i$, $n_i$, and $\mu_i$ denote the charge, carrier density, and mobility of the $i$-th carrier, respectively. 
Under the assumption of compensated carriers with opposite signs ($q_1 = -q_2 = e$) and equal densities ($n_1 = n_2 = n$), the expression for the Hall coefficient $R_1$ simplifies to
\begin{equation}
R_1 = \frac{1}{en} \frac{\mu_1 - \mu_2}{\mu_1 + \mu_2}.
\end{equation}
Evaluating the magnetic field dependence of $\rho_{yx}$, we estimated the carrier density and mobility for the dominant conduction channels. 
At ambient pressure, we obtained $n \approx 2 \times 10^{20}$~cm$^{-3}$ and $\mu \approx 400$~cm$^2$/Vs. 
As shown in the inset of Fig.~S1, these parameters exhibit only a weak pressure dependence in the low-pressure regime, consistent with the stability of the AFM phase.

\section{Computational Details and Band Warping in the AFM Phase}
First-principles density functional theory calculations were performed using the full-potential linearized augmented plane-wave method as implemented in the WIEN2k code~\cite{Blaha2020}.
The generalized gradient approximation proposed by Perdew, Burke, and Ernzerhof (PBE-GGA96) was employed for the exchange-correlation functional.
Spin--orbit coupling was included in all calculations.
The experimental lattice parameters reported by Tanida \textit{et al.}~\cite{Tanida2023} were used for the crystal structure.
The product of the radius of the muffin-tin sphere $R_{\rm MT}$ and the plane-wave cutoff $K_{\rm max}$, $R_{\rm MT}K_{\rm max}$ was set to 9.0.
The Brillouin zone integration was performed using a dense $k$-point mesh of $41 \times 41 \times 23$ (corresponding to 40,000 $k$-points in the full Brillouin zone). 
The self-consistent cycles were iterated until the convergence criteria of $10^{-6}$~Ry for the total energy and $10^{-5}$~$e$ for the charge distance were satisfied.

\begin{figure}[h]
\centering
\includegraphics[width=8cm,pagebox=cropbox,clip]{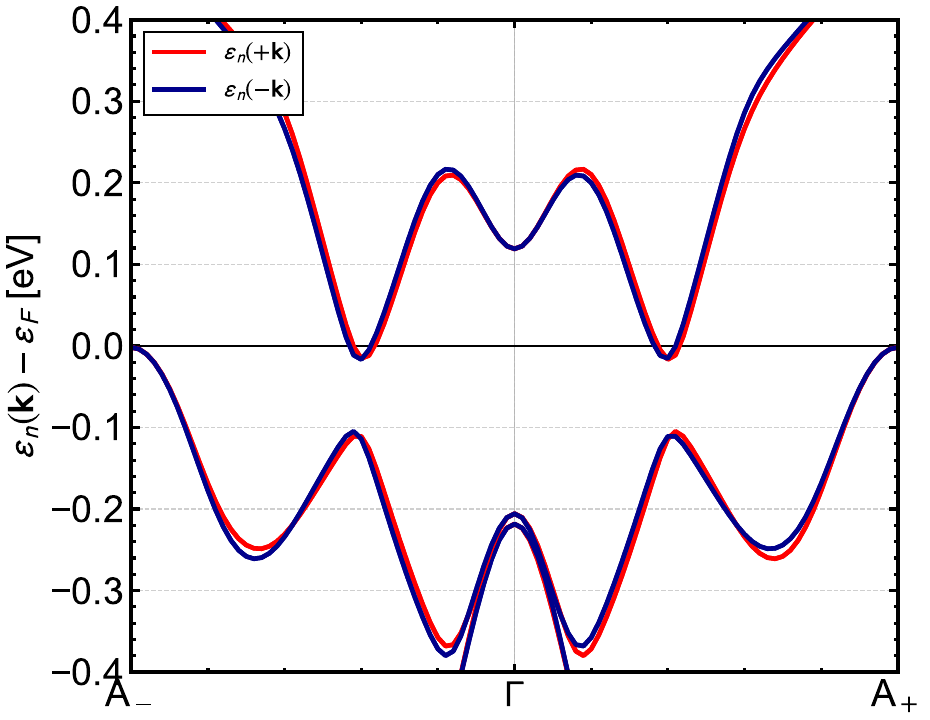}
\caption{Band asymmetry in the antiferromagnetic phase along the A$_{-}$-$\Gamma$-A$_{+}$ path with A$_{\pm}$=($\pm1/2$,$\pm1/2$,$\pm1/2$). Bands with inverted wavevectors are superimposed in red and blue to accentuate the asymmetry. These distinct bands correspond to different antiferromagnetic domain states.}
\label{SFig2}
\end{figure}

Unlike conventional AFM orders, a $\mathcal{PT}$-symmetric AFM state not only forbids breaking of the band degeneracy due to spin polarization but also induces asymmetric band warping along specific directions.
Indeed, our calculations reveal that, when the band dispersion is inverted and superimposed, $\varepsilon(\vec{k})$ and $\varepsilon(\vec{-k})$ do not coincide (Fig.~\ref{SFig2}).
This demonstrates that a $\mathcal{PT}$-symmetric electronic phase is realized, driven by the AFM order of Mn.
Because of magnetic symmetry constraints, such asymmetric warping occurs at wave vectors where $k_x$, $k_y$, and $k_z$ are all finite.

\end{document}